\documentclass[letterpaper,12pt]{article}   
\usepackage{osajnl2} 
\usepackage[draft]{hyperref} 
\usepackage{amsmath}
\usepackage{graphicx}
\usepackage{epstopdf}

\begin{document}

\title{Curvature sensors: noise and its propagation}

\author{Aglae Kellerer}
\address{Institute for Astronomy, \\ 640 N. A'ohoku Place, HI,
Hilo 96720, USA\\ kellerer@ifa.hawaii.edu}

\begin{abstract} 
The signal measured with a curvature sensor is here analyzed. 
In the outset, we derive the required minimum number of sensing elements at the pupil edges, in dependence on the total number of sensing elements. The distribution of the sensor signal is further characterized in terms of its mean, variance, kurtosis and skewness. It is established that while the approximation in terms of a gaussian distribution is correct down to fairly low photon numbers, much higher numbers are required to obtain meaningful sensor measurements for small wavefront distortions. Finally, we indicate a closed expression for the error propagation factor and for the photon-noise induced Strehl loss.
\end{abstract}

\ocis{110.0115, 110.1080} 

\maketitle 

\section{Introduction}

In an adaptive optics system, a wavefront sensor measures phase distortions in real-time. 
These measurements are used to update the shape of a deformable mirror located upstream in the optical beam. 
One such wavefront sensor was developed by F.\,Roddier and presented for the first time in 1988 \cite{roddier1988}. 
In a curvature sensor (C-S), two defocused images are recorded, one on each side of the nominal focal plane. The wavefront distortion is then inferred from the difference in intensity between the two images.
F.\,Roddier and his group at the Institute of Astronomy, Hawaii, implemented a low-order adaptive-optics (AO) system based on a curvature WFS with 13 actuators \cite{roddier1991}.  In the following years many other curvature based AO systems were successfully implemented \cite{PUEO}, \cite{NICI}, \cite{MACAO}, \cite{Subaru}. 

However, as adaptive optic systems evolve towards increasingly large numbers of sensing elements and actuators, curvature sensors tend to be discarded in favor of Shack-Hartmann (S-H) sensors.
This is mainly based on the assertion that the effect of measurement errors on the corrected phase increases steeply with the number of sensing elements in C-S. This steep, linear increase of the error propagation term was established by N. Roddier through simulations\cite{NicolasDiploma}. The author further reported on concurring analytic calculations by J.B. Shellan for more than 10 000 actuators. These calculations are however not publicly available. 
High-order curvature systems have been simulated by e.g. O.\,Lai et al.\,\cite{Lai} and Q.\,Yang et al.\,\cite{Yang}. Both authors conclude in general favor of high-order C-S systems. Yang et al. conclude: ``No degradation of the phase transfer function with increasing actuator number is observed, implying that still higher actuator number systems are possible with no significant modification of the basic concept and design''. 
We feel that, in addition to these results, it is necessary to analyze the effect of measurement noise on the performance of curvature systems, and in particular on the performance of high-order systems.
    
Sec.\,\ref{sec:2} of this article considers the special role of the sensing elements that lie on the pupil edge. The minimum number of such edge elements is derived in dependence on the total number of sensing elements. This result is then used in section\,\ref{sec:4} where C-S with up to 225 elements are simulated.  
Sec.\,\ref{sec:3} analyzes the measurement noise in curvature sensors: the quantity  measured by curvature sensors is $v=(x-y)/(x+y)$, where the photon numbers $x, y$ follow a Poisson distribution. Can the distribution of $v$ be approximated by a Gaussian and is the 68\,\%-confidence interval well approximated by $\overline{v}\pm\sigma_v$? Sec.\,\ref{sec:3} further derives the number of photons needed to obtain meaningful measurements of $v$.
Sec.\,\ref{sec:4} studies the effect of measurement noise on the phase reconstruction. We agree with N. Roddier's result, that were obtained for a fixed size of the sensing elements and a fixed defocus length. However, for a constant pupil surface -- hence a decreasing sensing-element size -- and with an adjusted extrafocal distance, the error propagation factor is found to be almost independent of the number of sensing elements. Sec.\,\ref{sec:4}  concludes with a closed expression for the error propagation factor and an estimate of the Strehl decrease due to photon noise.

\section{Minimum number of edge sensors}\label{sec:2}
\subsection{Signal measured by a curvature sensor}

The output signal of a curvature wavefront sensor is the intensity difference $v=(x-y)/(x+y)$,  where $x$ and $y$ are the photon numbers measured on either side of the focal plane.
In closed-loop AO observations, $v$ tends towards 0 and can therefore be expressed as a function of the wavefront phase, $\phi$: \cite{NOAO}
\begin{equation}\label{eq:Roddier}
v(\vec r) = \frac{-\lambda f (f-l)}{2 \pi l} [ \vec{\nabla}(P(\frac{f}{l} \vec r)) \cdot \vec{\nabla}(\phi(\frac{f}{l} \vec r)) + P(\frac{f}{l} \vec r) \cdot \nabla ^2 \phi(\frac{f}{l} \vec r)]
\end{equation}
$\vec r$ is the position vector in the detector plane. 
$f$ is the focal length, $l$ the distance between the detector and the focal plane and $f/l\,\vec r$ is thus the position vector in the pupil plane. $\lambda$ is the wavelength and $P$ -- the pupil transmission function -- equals 1 inside and 0 outside the pupil.
Eq.\,\ref{eq:Roddier} makes use of the Fresnel approximation: $D^2/((f-l)\lambda) \geq 1$, where $D$ is the pupil diameter. Eq.\,\ref{eq:Roddier}  further assumes the size of the diffraction patterns to be negligible compared to the characteristic scales of intensity fluctuations in the detector plane. 
This sets a condition on the extra focal distance: $\lambda f^2 << r_0^2 \, l$, where $r_0$ is the Fried parameter.
In regions where the pupil transmission is uniform, the measured signal is the laplacian of the wavefront phase.
On the pupil edge -- i.a. along the central obstruction and along the telescope spiders -- curvature sensors measure curvature and radial tilt. 
Phase aberrations with a laplacian equal to zero are sensed solely at the pupil edges. 

\subsection{Zero-laplacian Zernike modes}
The phase of the distorted wavefront is expressed in terms of a series of Zernike polynomials: 
\begin{equation}\label{eq:zernike}
\phi =\sum_{n \geq 0} \sum_{ m=-n} ^{n} a_n^m\,Z_n^m
\end{equation}
$n$ and $m$ are integers and $n-|m|$ is even. $a_n^m$ is the coefficient of the Zernike polynomial $Z_n^m$. In atmospherically distorted wavefronts, the lowest order Zernike modes have the highest variance and need to be corrected by the AO system first. The first order Zernike mode, piston, is an exception: it corresponds to a constant phase change over the whole pupil and does not affect the image quality.
Table \ref{tab:laplacians} lists the laplacian of the fifteen lowest Zernike polynomials.
Eight out of fourteen polynomials (piston excluded) have zero laplacian, and their contribution to the distorted wavefront is therefore only measured at the pupil edges. 
In an ideal AO system $K+1$ `sensor-actuator' pairs (i.e. $K+1$ sub-apertures) are required to sense and correct $K$ modes of the wavefront phase.
It is assumed that the lowest Zernike modes are corrected first (piston excluded). $K$ is the number of corrected modes and $n$ the radial index of the highest, corrected mode.  
Since there are $p+1$ Zernike polynomials for each radial index $p$, $K$ and $n$ are related by the following inequality:
\begin{eqnarray}\label{eq:radial}
\frac{n (n+1)}{2}-1 < K \leq \frac{(n+1)(n+2)}{2} -1
\end{eqnarray}

\begin{table}[htdp]
\caption{Laplacian of Zernike polynomials: 
$\nabla^2 Z(r,\theta) = \frac{1}{r} \frac{\partial}{\partial r}(r \frac{\partial Z}{\partial r}) + \frac{1}{r^2} \frac{\partial^2 Z}{\partial \theta^2}$}
\begin{center}
\resizebox{\textwidth}{!}{
\begin{tabular}{ l l l l l l}  \hline \hline
n & m  & & & & \\ 
  & 0 & 1 & 2 & 3 & 4  \\ \hline
0 & $\nabla^2Z_0^0$=0 & & & & \\

1 & & $\nabla^2Z_1^{-1}= 0$   & & & \\
  & & $\nabla^2 Z_1^{1}= 0$   & & & \\
  & & tip \& tilt   & & & \\

2 & $\nabla^2 Z_2^0= 8\sqrt{3} $ & & $\nabla^2 Z_2^{-2}= 0$   & & \\
 & defocus & & $\nabla^2 Z_2^{2}= 0$   & & \\
 &  & & astigmatism (3rd order)   & & \\

3 &  &  $\nabla^2 Z_3^{-1}= 24 \sqrt{8} r \sin(\theta)$ &  &  $\nabla^2 Z_3^{-3}= 0$ & \\
 &  &  $\nabla^2 Z_3^{1}= 24 \sqrt{8} r \cos(\theta)$   & & $\nabla^2 Z_3^{3}= 0$ & \\
 &  &  coma  & &trefoil & \\

4 & $\nabla^2 Z_4^{0}=24 \sqrt{5} (4r^2 - 1)$ &   &  $\nabla^2 Z_4^{-2}= 120 \sqrt{10} r^2 \cos(2 \theta)$ & & $\nabla^2 Z_4^{-4}= 0$\\
  &  spherical &   &  $\nabla^2 Z_4^{2}= 120 \sqrt{10} r^2 \sin(2 \theta)$ & & $\nabla^2 Z_4^{4}= 0$\\
 & &  & astigmatism (5th order) & & ashtray \\

\end{tabular}}
\end{center}
\label{tab:laplacians}
\end{table}%

\begin{figure}[htbp]
\begin{center}
\includegraphics[width=0.4\textwidth]{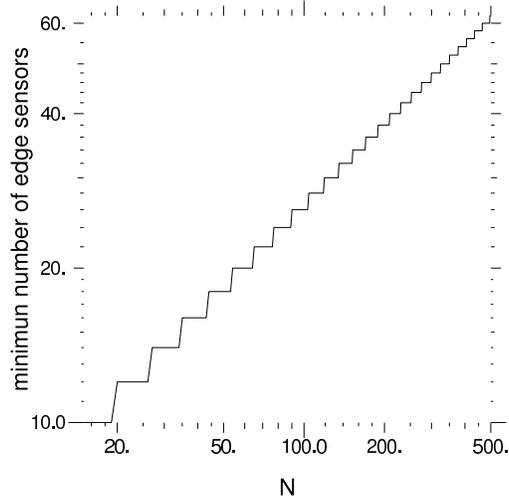}
\caption{Minimum number of sensing elements at the aperture edge required to sense the $N$ lowest Zernike modes.}
\label{fig:edge}
\end{center}
\end{figure}

The Zernike modes for which the absolute value of the meridional index, $m$, equals the radial index, $n$, have zero laplacian. 
There are, consequently, two zero-laplacian polynomials per radial index, hence in order to correct the first $K$ Zernike polynomials, a curvature sensor needs at least $N_e=2\,n$ sensors at the aperture edge, where $n$ is determined by Eq.\,\ref{eq:radial}:
\begin{eqnarray}
N_e = 2 \cdot \left\lfloor \frac{1}{2}(\sqrt{8K+9} -1) \right\rfloor
\label{eq:min}
\end{eqnarray}
$\lfloor\,\rfloor$ is the floor function.
Figure\,\ref{fig:edge} traces the minimum number of edge sensors in dependence on the number of modes to be corrected.
Instead of increasing the number of edge sensors, it is also possible to smoothen the pupil edges. This distributes the radial tilt signal over more sensors, at the expense of smaller signal in each individual sensor.

\section{Uncertainty of the sensor measurement}\label{sec:3}

In this section, a single sub-aperture of the sensor is considered. 
Be $x$ and $y$ the photon numbers measured on either side of the focal plane.  
Be $X$ and $Y$ their expected values, i.e. the number of photons that one would measure in average under frozen atmospheric conditions and repeated measurements. The difference between $x, y$ and $X, Y$ is taken to be solely due to photon noise. This is a valid approximation if the sky background signal is low and if the detector readout noise and dark current are negligible. Existing curvature sensors use avalanche photo-diodes where the readout noise and dark current are indeed negligible.

The variable $v=(x-y)/(x+y)$ serves as estimate of the deviation $V=(X-Y)/(X+Y)$. $x$ and $y$ are independent Poisson distributed random variables with mean value $X$ and $Y$.  To judge the precision of the estimate the parameters of the distribution of $v$ need to be known, i.e. the mean, the variance, the skewness and the kurtosis. These are closely related to the moments or, more specifically, the cumulants, $\kappa_i$, of the distribution.
In closed-loop AO observations, $V$ tends towards 0. Furthermore the photon numbers, $Z=X+Y$, are usually large. The parameters of the distribution of $v$ are then, as will be seen, simple: the mean and the skewness are zero and both the variance and the excess kurtosis are $1/Z$. It is nevertheless of interest to derive the exact solution for the more general case where $V$ is not zero, and $Z$ is not very large. 

\subsection{Non-central moments of $x/z$}

It is convenient to deal first with the variable $x/z$, rather than $v=2x/z-1$. The cumulants, $\kappa_i$ of $v$ are subsequently expressed in terms of the cumulants, $k_i$, of $x/z$:
\begin{eqnarray}
\kappa_1&=& 2\,k_1-1\\\label{eq:cum1}
\kappa_2&=& 4\,k_2\\\label{eq:cum2}
\kappa_3&=& 8\,k_3\\\label{eq:cum3}
\kappa_4&=& 16\,k_4\label{eq:cum4}
\end{eqnarray}
The joint distribution of $x$ and $z$ equals the product of the Poisson probability, $p(z; Z)$, for the variable $z$ with mean value $Z$, and the binomial probability, $b(x; z,p)$, that out of the z events $x$ are of the type that occurs with relative probability $p=X/Z$:
\begin{eqnarray}
f(x, z) = p(z, Z) \; b(x; z,p)
\end{eqnarray}
The outcome $z=0$ is excluded since $v$ is then undefined. The probabilities $p(z; Z)$ are thus normalized to the positive integers $z$, i.e. are increased by the factor $1/(1-\exp(-z))$ relative to the familiar Poisson probabilities.

The $i$-th moment, $u_i=<(x/z)^i>$ can be written in the form:
\begin{eqnarray}
<(x/z)^i> &=& \sum_{z>0} ( \sum_{x = 0}^{ z} x^i  \; b(x; z,p)) z^{-i} \; \frac{ p(z, Z) }{1-e^{-z}}\\
&=& \sum_{z>0} \mu_i(z,p) \; z^{-i} \;  \frac{ p(z, Z) }{1-e^{-z}} \label{eq:mom}
\end{eqnarray}
where $\mu_i(z,p)$ are the moments about zero of the binomial distribution with mean value $z$ and parameter $p$.
Using the familiar equations for these moments one obtains the expressions:
\begin{eqnarray}
\mu_1 / z &=& p\\\label{eq:m1}
\mu_2 / z^2 &=& (p-p^2)/z + p^2\\\label{eq:m2}
\mu_3 / z^3 &=& (p-3\,p^2 + 2\,p^3)/z^2 + 3\, (p^2 - p^3)/ z + p^3\\\label{eq:m3}
\mu_4 / z^4 &=& (p-7\,p^2 + 12\,p^3-6\,p^4)/z^3 +  (7\,p^2 - 18\,p^3+11\,p^4)/ z^2 +6\,(p^3-p^4)/z + p^4\label{eq:m4}
\end{eqnarray}
The moments $u_i=<(x/z)^i>$ of the variable $x/z$ are obtained by summation over $z$ according to Eq.\,\ref{eq:mom}. The problem is thus reduced to the determination of the Poisson mean values $<z^{-i}>$. Since there are no closed expressions, one needs to evaluate those mean values numerically, which is conveniently done in term of the ratios:
\begin{eqnarray}
\rho_i &=& <z^{-i}>  / Z^{-i} \\
&=& Z ^i \;\sum_{z>0} z^{-i} \;  \frac{ p(z, Z) }{1-e^{-z}} \\
&=& Z ^i \;\sum_{z>0} z^{-i} \; \frac{e^{-Z}\cdot Z^z }{z!} \cdot \frac{1}{1-e^{-z}} \label{eq:rho}
\end{eqnarray}
The $\rho_i$ converge to unity at large $Z$. 

\subsection{Transition to the cumulants}

The cumulants of the distribution of $x/z$ stand in a general relation to the moments:
\begin{eqnarray}
k_1&=& u_1\\
k_2&=& u_2-u_1^2\\
k_3&=& u_3-3\,u_2 \,u_1 + 2\,u_1^3\\
k_4&=& u_4-4\,u_3 \,u_1 + 6\,u_2\,u_1^2 -3\,u_1^4 - u_3 + 3\,u_2\,u_1 - 2\,u_1^3\\
\end{eqnarray}
Using Eqs.\,\ref{eq:m1}-\ref{eq:rho} and switching in terms of Eqs.\,\ref{eq:cum1}-\ref{eq:cum4} to the cumulants of $v$ one obtains:
\begin{eqnarray}
\kappa_1 &=& 2\,p-1  \\
\kappa_2&=&4 \; (p-p^2) \rho_1 / Z \\
\kappa_3 &=& 8 \; (p-3\,p^2 +2\,p^3) \rho_2 / Z^2\\
\kappa_4 &=& 16\,(p-7\,p^2+12\,p^3-6\,p^4)\,Z^2+16 \; (3p^2-6\,p^3 +3\,p^4) \,(\rho_2-\rho_1^2) / Z^2
\end{eqnarray}
Instead of the parameter $p=X/Z$, i.e. the mean value of $x/z$, one wishes to use the parameter $V=(X-Y)/(X+Y)$: 
\begin{eqnarray}
\kappa_1 &=& V \\ 
\kappa_2 &=& (1-V^2) \; \rho_1 / Z\\
\kappa_3&=& -2\,V\,(1-V^2)\,\rho_2/Z^2\\
\kappa_4&=& (-2+8\,V^2-6\,V^4)\,\rho_3/Z^3 + 3(1-V^2)^2(\rho_2-\rho_1^2)/Z^2
\end{eqnarray}
The commonly used parameters of the distribution of $v$, i.e. the mean value $<v>=\kappa_1$, the variance, $\sigma^2_v=\kappa_2$, the skewness, sk$_v=\kappa_3/\kappa_2^{1.5}$ and the (excess) kurtosis, kur$_v=\kappa_4/\kappa_2^2$ are, therefore:
\begin{eqnarray}
<v> &=& V \\ 
\sigma^2_v &=& (1-V^2) \; \rho_1 / Z\\
\rm{sk}_v &=& -\frac{2\,V}{\sqrt{1-V^2}}\,\frac{\rho_2}{\rho_1^{1.5}}\frac{1}{\sqrt{Z}}\\
\rm{kur}_v &=& (\frac{4\,V^2}{1-V^2}-2)\frac{\rho_3}{\rho_1^2}\frac{1}{Z} + 3\,(\frac{\rho_2}{\rho_1^2}-1)
\end{eqnarray}
For large $Z$, $\rho_1$ tends to unity, and, as can be shown, $(\rho_2/\rho_1^2-1)$ converges to $1/Z$. The formulae then reduce to:
\begin{eqnarray}\label{eq:s1}
\sigma^2_v &=& (1-V^2)  / Z\\\label{eq:s2}
\rm{sk}_v &=& -\frac{2\,V}{\sqrt{1-V^2}}\frac{1}{\sqrt{Z}} \\\label{eq:s3}
\rm{kur}_v &=& (\frac{4\,V^2}{1-V^2}+1)\frac{1}{Z} \label{eq:s4}
\end{eqnarray}
As stated in the outset, the parameters become very simple for small $V$, i.e. for nearly equal light intensities in the two images, and if $Z$ is large. The mean and skewness of $v$ are then zero and the variance and the (excess) kurtosis are $1/Z$.

\subsection{Visualisation}

\begin{figure}[htbp]
\begin{center}
\includegraphics[width=0.4\textwidth]{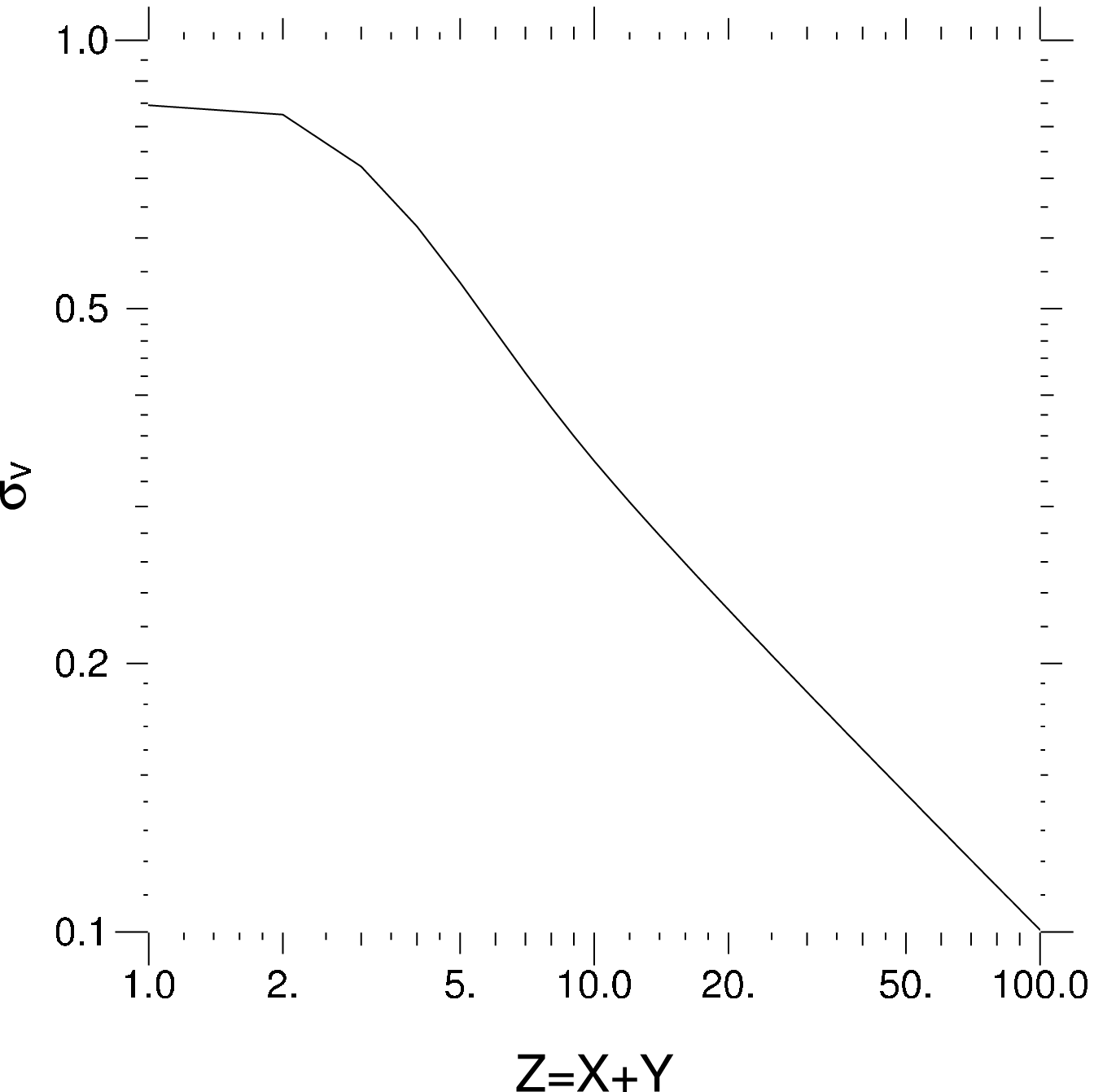}
\includegraphics[width=0.4\textwidth]{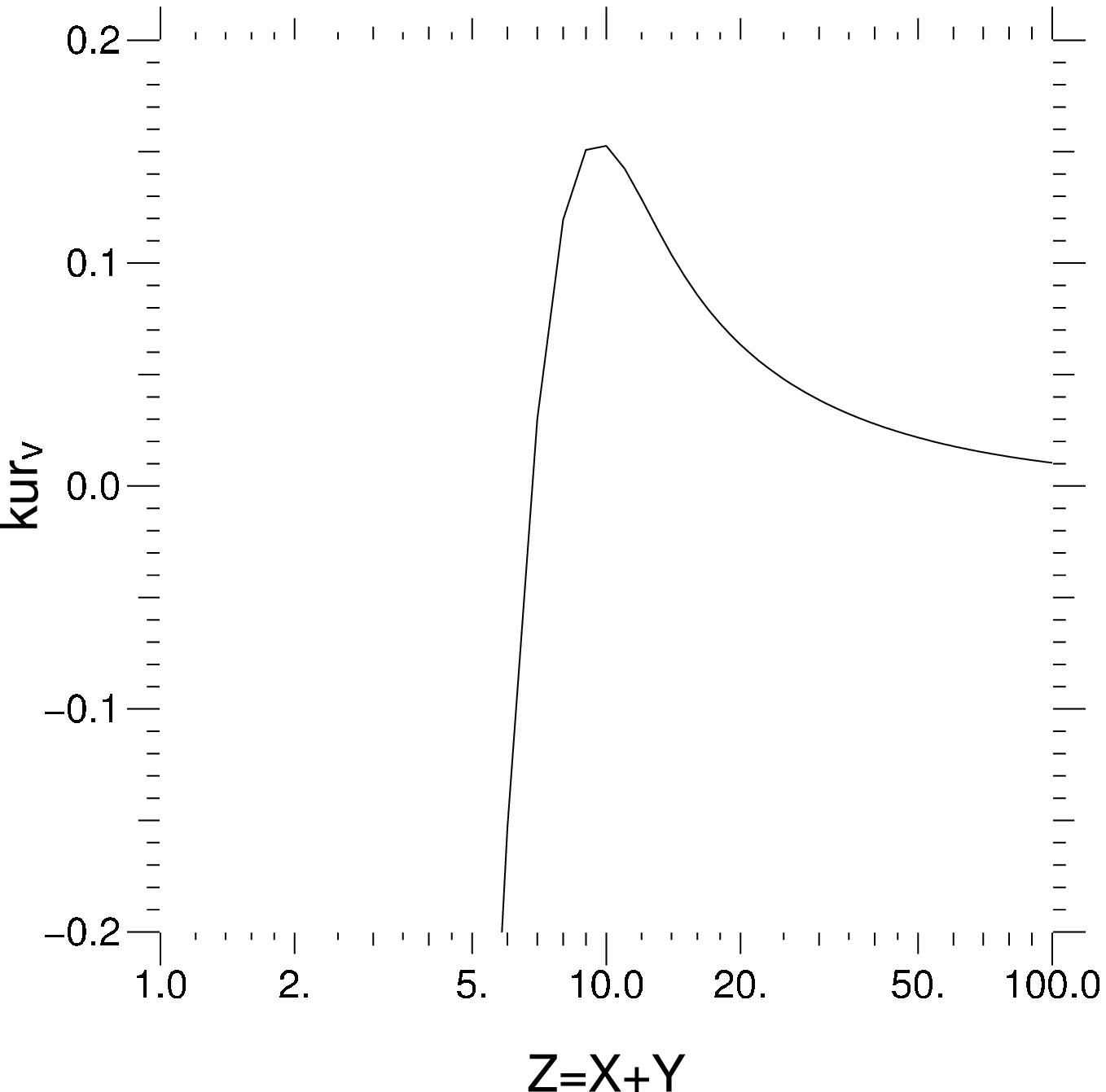}
\caption{The standard deviation and the kurtosis of $v$ in dependence on $Z$ for $V=0$:
The approximation in terms of the Gaussian distribution is justified down to fairly low photon values ({\it right panel\/}), but much higher photon numbers are required to obtain meaningful measurements at small values of $V$, i.e. for small wavefront distortions ({\it left panel\/}).}
\label{fig:a1}
\end{center}
\end{figure}

Fig.\,\ref{fig:a1} gives the standard deviation and the excess kurtosis of the measured $v$ for the case of equal light intensity of the two AO images, i.e. for $V=0$. It shows that the kurtosis is close to zero even for fairly low photon numbers, $Z$. The approximation in terms of the Gaussian distribution is, thus, justified and the standard error range in terms of the 68\,\%-confidence range for a measured value $v$ can be set equal to $v\,\pm\,\sigma$ while the 95\,\%-confidence range is $v\,\pm\,2\,\sigma$. Since $\sigma$ still has the substantial value 0.1 for $Z=100$, one requires much higher photon numbers to obtain meaningful measurements at small values of $V$. Even at the large photon number $Z=10^4$ the standard error of $v$ is 0.01, i.e. one needs a measured value $v\geq0.02$ to reject the null-hypothesis of equal light intensity in the two images ($V=0$) on the 95\,\%-confidence level. This is not trivial, since the measured $v$-values tend to be small in curvature-sensor systems for AO.

\begin{figure}[htbp]
\begin{center}
\includegraphics[width=0.4\textwidth]{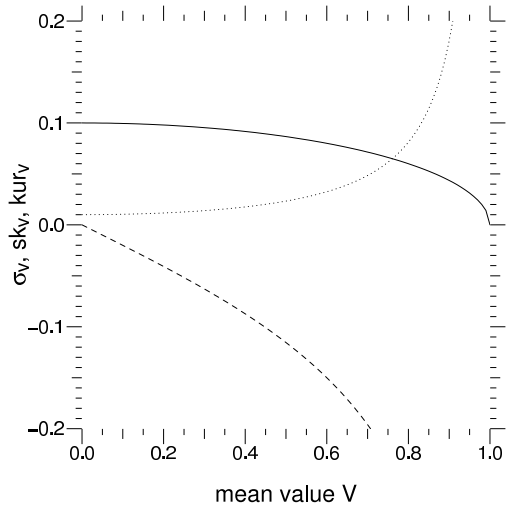}
\caption{The standard deviation (solid line), the skewness (dashed) and the kurtosis (dotted) of $v$ in dependence on $V$ for $Z=100$. 
The decrease of the standard error is not substantial in the [0,0.2] range representative of closed-loop adaptive optics observations.  At the large photon numbers that are thus required, the Gaussian approximation  remains valid despite the increase of the skewness and kurtosis.}
\label{fig:a2}
\end{center}
\end{figure}

This being a critical issue, it is useful to consider the distribution parameters not just for $V=0$. While fairly low values of $V$ are important in practice, it is instructive to take a broader view and consider the full range of possible intensity differences. Fig.\,\ref{fig:a2} gives for $Z=100$ the standard error, the skewness and the kurtosis in dependence on $V$. As seen in Eqs.\,\ref{eq:s2}-\ref{eq:s4}, the shape of these dependencies remains the same for higher photon numbers, $Z$, while the absolute values of the standard deviation and the skewness decrease proportionally to $1/\sqrt{Z}$ and the kurtosis decreases proportionally to $1/Z$. At the large photon numbers $Z$ that are required the Gaussian approximation remains thus valid inspite of the increase of the absolute value of the skewness and the kurtosis at large intensity differences. The decrease of the standard error is helpful in principle, but is not substantial at moderate values of $V$.

\begin{figure}[htbp]
\begin{center}
\includegraphics[width=0.4\textwidth]{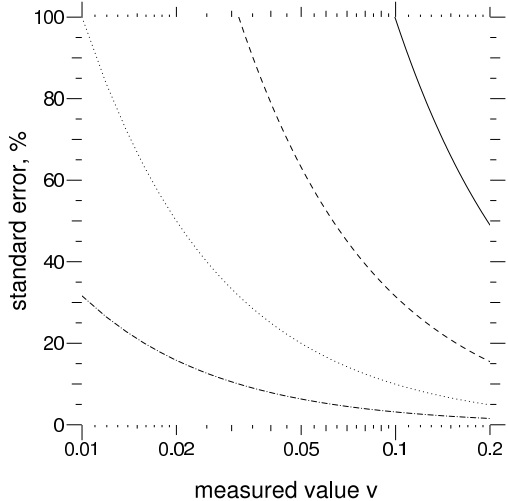}
\caption{The need for large photon numbers: percent standard error in dependence on the measured value $v$ for photon numbers $Z=X+Y=$100 (solid line), $10^3$ (dashed), $10^4$ (dotted) and $10^5$ (dashed-dotted).}
\label{fig:a3}
\end{center}
\end{figure}

Fig.\,\ref{fig:a3} illustrates the need for large photon numbers directly. It gives for any measured value $v$ and for different photon numbers the 68\,\%-confidence range of the estimated value $V$.

\section{Error propagation}\label{sec:4}

\subsection{Definition and context}

An error on the sensor measurement translates into an error on the reconstructed or corrected wavefront: 
Inside a correction loop, the error on the sensor measurement propagates on the voltages  through the {\it command matrix\/}, onto the mirror surface through the actuator {\it influence functions\/} and finally to the phase as the wavefront reflects on the mirror.
If the sensor is not part of an AO control-loop, the error propagates through a phase reconstruction algorithm. 
The error propagation factor equals the phase variance due to statistically independent random measurements of unit variance. It is defined by the following relation.
\begin{eqnarray} \label{eq:G}
G=\frac{\sigma_\phi^2}{\sigma_v^2}
\end{eqnarray}
where $\sigma_v^2$ is the variance of the sensor measurements and $\sigma_\phi^2$ the induced variance of the corrected or reconstructed phase.

The error propagation factor in Shack-Hartmann sensors has been analyzed by Hudgin \cite{hudgin1977}  and Fried  \cite{fried1977}.
Both, Hudgin's and Fried's wavefront-reconstruction algorithms lead to a logarithmic increase of the error propagation term with the number of sensing elements: $G\propto \ln (N)$. Both authors assume that the size of the sampling elements is constant with changing $N$. 

\subsection{Curvature sensors}\label{sec:curvprop}

Simulations by N. Roddier for actuator numbers in the 5-100 range have shown that the error propagation term in curvature sensors increases linearly with $N$ \cite{NicolasDiploma}. 
The behavior of the error propagation term is here analyzed. It is found -- in agreement with N. Roddier's simulations -- that $G$ increases linearly with the number of sub-apertures, if the pupil size increases with $N$ and if the defocus distance is kept fixed. The behavior of the error propagation term for a given pupil surface and an increasing defocus distance is then examined and a closed expression for the error propagation factor is indicated.
 
 \subsubsection{Phase reconstruction}
 
To obtain an expression for the error propagation factor in curvature sensors, we adapt Hudgin's phase-reconstruction algorithm. Let us briefly recall the principle of this algorithm\,\cite{hudgin1977}: 
A S-H sensor  determines the phase difference, e.g. d$\phi_x(l, n) = \phi(l+1, n) - \phi(l, n)$ and d$\phi_y(l, n) = \phi(l, n+1) - \phi(l, n)$.  
 For a set of noiseless slope measurements, the phase value at a grid point $(i,k)$ relates to the phase values of its four nearest neighbors :
\begin{eqnarray}
\phi(l,n)  =  ( \phi(l-1,n) +  {\rm d}\phi_x(l-1,n)+  \phi(l+1,n) - {\rm d}\phi_x(l,n) + \nonumber \\ 
\phi(l,n-1) +{\rm d}\phi_y(l,n-1) + \phi(l,n+1)  -{\rm d}\phi_y(l,n) ) / 4   \label{eq:p2}
\end{eqnarray}
This relation only applies to inner grid points and Hudgin did consider an infinitely large pupil. For corner (resp. border) points we modify this relation to use only the two (resp. three) nearest neighbors. The denominator in Eq.\,\ref{eq:p2} is accordingly changed to 2 (resp. 3). 

For slope values with measurement errors there is no perfect fit that fulfills this condition, but the best fit minimizes the mean squared deviations from the equality. 
Hudgin referred, accordingly, to a reconstruction method where one starts with an arbitrary phase estimate, $\phi(i,k)$. In each iteration cycle the phase estimates on each grid point are replaced by the average of the phase values of its four closest neighbors, each adjusted by the fixed values of the phase differences:
\begin{eqnarray}
\phi(l,n) \leftarrow \; ( \phi(l-1,n) +  {\rm d}\phi_x(l-1,n)+  \phi(l+1,n) - {\rm d}\phi_x(l,n) + \nonumber \\ 
\phi(l,n-1) +{\rm d}\phi_y(l,n-1) + \phi(l,n+1)  -{\rm d}\phi_y(l,n) ) / 4
\label{eq:p3}
\end{eqnarray}

The new value $\phi(l,n)$ is the one that best fits  the 4 latest phase estimates for the nearest neighbors and the values of the associated slopes.
Eq.\,\ref{eq:p3} suggests a simplification: The sum of the 4 phase-difference terms in the equation is an estimate of the second-order phase-difference at grid point $(l,n)$:
\begin{eqnarray}
{\rm d}^2\phi(l,n) = ( {\rm d}\phi_x(l-1,n) -  {\rm d}\phi_x(l,n)   +  {\rm d}\phi_y(l,n-1) - {\rm d}\phi_y(l,n) ) / 4
\label{eq:p4}
\end{eqnarray}
And Eq.\,\ref{eq:p3} can, accordingly, be written in the simplified form:
\begin{eqnarray}
\phi(l,n) \leftarrow \; ( \phi(l-1,n) +  \phi(l+1,n)  + \phi(l,n-1) + \phi(l,n+1)  ) / 4 + {\rm d}^2\phi(l,n)
\label{eq:p5}
\end{eqnarray}
with the modification that has been noted for the border and corner points.

For a set of $N$ sensor measurements $v(l,n)$ ($l,n=1..\sqrt{N}$) the second-order phase-differences are obtained through: 
\begin{eqnarray} 
{\rm d}^2\phi(l,n)  = \nabla^2 \phi (l,n)\cdot a^2/4 = -\frac{\pi\,l\,a^2}{2\,\lambda\,f\,(f-l)}\cdot v(l,n) 
\label{eq:vtoc}
\end{eqnarray}
where $a$ is the distance between two adjacent grid points. 
The $N$ phase values are obtained via the iteration method described by Eq.\,\ref{eq:p5}.

\subsection{Sensing elements of constant surface and fixed extra-focal distance}\label{sec:nicolas}

The error propagation factor is computed for the parameter values listed in the first column of Table\,\ref{tab:par1}: a fixed sub-aperture size $a$ and extra-focal distance $l$. In agreement with N.\,Roddier's\,\cite{NicolasDiploma}, $G$ increases linearly with the number of sensing elements:
\begin{eqnarray}
G= (2.20\pm0.05)\,10^3\cdot N - (2.03\pm0.05)\,10^4
\label{eq:linearG}
\end{eqnarray}

\begin{table}[h]
\caption{Parameter values for the computation of the error propagation term.}
\label{tab:par1}
\begin{center}       
\begin{tabular}{ l | l l l}  \hline 
& Sec.\,\ref{sec:nicolas} & Sec.\,\ref{sec:constant} & Sec.\,\ref{sec:yao}\\ \hline
Pupil shape: & square & square & circular\\ 
Pupil size, $D$: & $a\cdot\sqrt{N}$ & 8\,m & 3\,m\\ 
Sub-aperture size, $a$: & 1\,m & $D/\sqrt{N}$ & $D/\sqrt{N}$\\
Extra-focal distance, $l$: & 0.5\,m & 0.05\,$\sqrt{N}$\,[m] & 0.8\,$\sqrt{N}$\,[m] \\
Wavelength, $\lambda$ & $0.7\,\mu$m &$0.7\,\mu$m &$0.7\,\mu$m\\
focal length, $f$: & 120 & 120 & 180\\
Number of sensing elements, $N$: & 25..225 & 25..225 & 25..225\\
\hline
\end{tabular}
\end{center}
\end{table} 

\subsection{Fixed pupil surface and adjusted extra-focal distance}\label{sec:constant}
On a given telescope and for a constant pupil surface, the sub-aperture size decreases with the number of sensing elements: $a=D/\sqrt{N}$. 
The measurements would eventually get blurred by diffraction if the extra-focal distance was kept constant, because the diffraction pattern has a fixed size in the image plane: $\lambda f/r_0$, where $r_0$ is the Fried parameter. 
The extra-focal distance therefore needs to be increased, $l\propto \sqrt{N}$, so that the size of the sensing elements in the imaging plane, $a \,l/f$, stay constant. 

\begin{figure}[htbp]
\begin{center}
\includegraphics[width=0.4\textwidth]{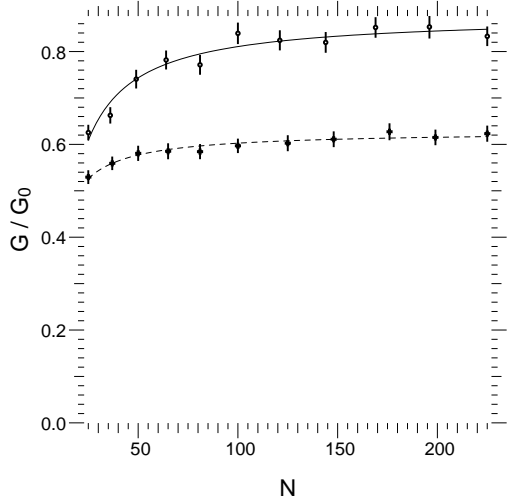}
\caption{Error propagation term, $G$, in dependence on the number of sub-apertures. {\it Full line\/}:  in terms of the iterative method for the parameters values  listed in the second column of Table\,\ref{tab:par1} (square pupil). {\it Dotted line\/}: in terms of the adaptive-optics simulation package {\it yao\/} for the parameter values listed in the third column of Table\,\ref{tab:par1} (circular pupil). $G_0$ is defined by Eq.\,\ref{eq:G0}.}
\label{fig:G}
\end{center}
\end{figure}

The error propagation factor is thus computed for the parameter values listed in the second column of Table\,\ref{tab:par1}: a constant pupil surface and an adjusted extra-focal distance. As seen on Fig.\,\ref{fig:G}, $G$ is then almost independent of the number of sensing elements. 
A general expression for the error propagation factor is deduced from these numerical results, using Eq.\,\ref{eq:vtoc}:
\begin{eqnarray}
G = \frac{\sigma^2_{\phi} }{\sigma_v^2} = G_0\cdot \left[ (0.85\pm 0.05) - \frac{(6.0 \pm 0.5)}{N}\right] \\
G_0= (\frac{l}{\sqrt{N}})^2\cdot(\frac{D}{f})^4 \cdot \frac{1}{\lambda^2}
\label{eq:G0}
\end{eqnarray}
This equation translates that the error propagation factor increases linearly with $N$ for a unit sampling size and a constant extra-focal distance (Eq.\,\ref{eq:linearG}), and that it is almost independent of $N$ when the pupil surface is constant and the extra-focal distance is adjusted (see Fig.\,\ref{fig:G}).

\subsection{Comparison  against an adaptive-optics simulation package}\label{sec:yao}
\begin{figure}[htbp]
\begin{center}
\includegraphics[width=0.4\textwidth]{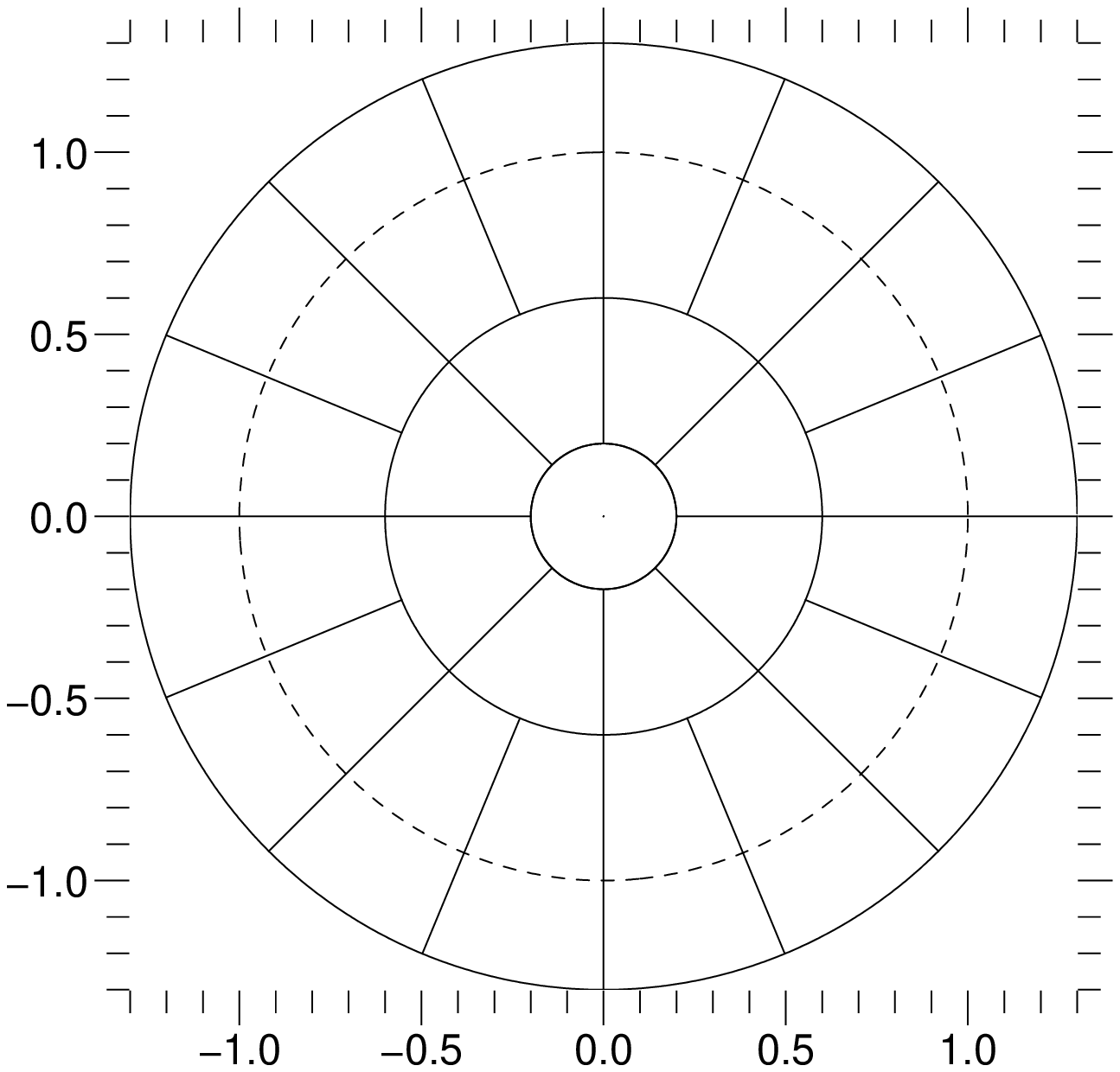}
\includegraphics[width=0.4\textwidth]{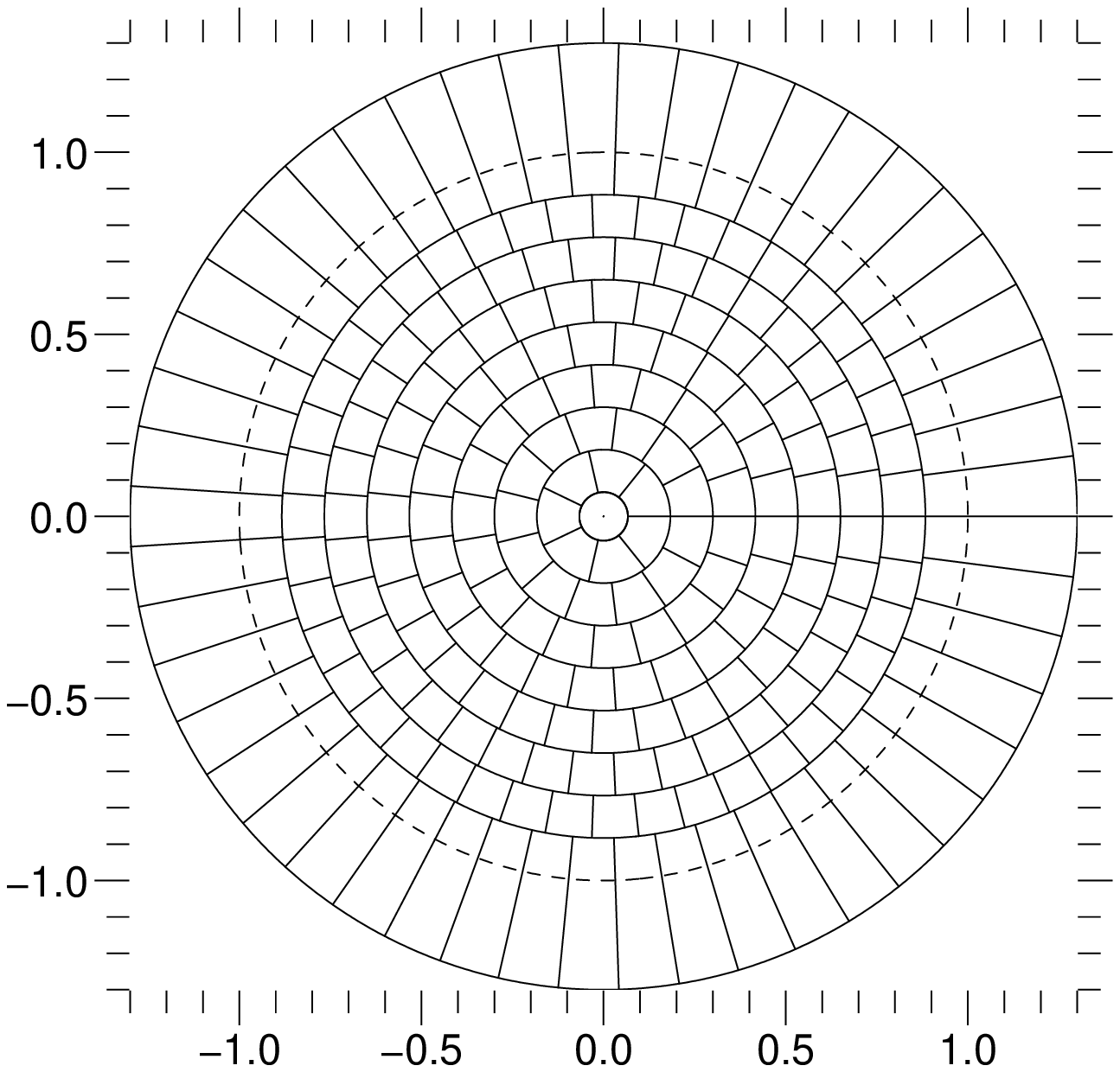}
\caption{Configurations of two of the simulated sensors: the wavefront is sampled over 25 and 225 elements. {\it Dashed line\/}: Pupil edge.}
\label{fig:configurations}
\end{center}
\end{figure}

In this section, the values obtained in sec.\,\ref{sec:nicolas} and \ref{sec:constant} are checked against those obtained with the adaptive-optics simulation package {\it Yao\/}\,\cite{yao}. 
In sections\,\ref{sec:nicolas} and \ref{sec:constant} the error propagation factor has been calculated on square grids, while with {\it Yao\/} the pupil is circular. 
The sampling geometry is inspired by Roddier et al.  \cite{roddier1991}: the sub-apertures are distributed along concentric rings and cover equal surfaces.  The outer elements have the highest error propagation factors and it is therefore desirable to minimize their number\footnote{Without any sensing element on the pupil edge, the error propagation factor is reduced by a factor of about 5.}: we choose the number of outer elements to be above the minimum required number (Eq.\,\ref{eq:min}) by 10-20\%.
Figure\,\ref{fig:configurations} shows two of the eleven generated sensor configurations.

The procedure adopted to calculate the error propagation factor under {\it Yao\/} is the following:
\begin{itemize}
\item[--] given $N$ sub-apertures, {\it Yao\/} computes $N$ {\it influence functions\/}: $N$ wavefronts resulting from unit voltages applied  to actuators $i=1... N$.
\item[--] the {\it command matrix\/}, which converts a set of $N$ sensor measurements into $N$ voltage values, is determined from its inverse, the {\it interaction matrix\/}. 
\item[--] for statistically independent random measurements of standard deviation $\sigma_v=0.1$, the set of corresponding voltage values is obtained through the command matrix. The wavefront is the sum of the $N$ influence functions weighted by the $N$ voltages.  It is expressed in radians.
\item[--]  $\sigma_{\phi}^2(i)$ is the phase variance of the $i$-th wavefront realization. The error propagation factor is determined from the following equation:
\begin{eqnarray}
G= \frac{1}{1000} \sum_{i=1}^{1000} \sigma_{\phi}^2(i) / \sigma_v^2
\end{eqnarray}
\end{itemize}

\begin{figure}[htbp]
\begin{center}
\includegraphics[width=0.4\textwidth]{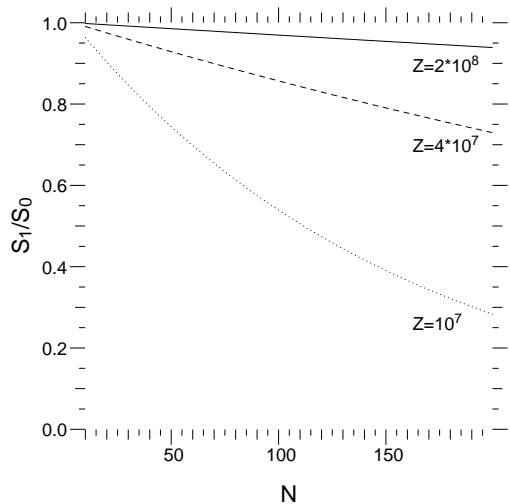}
\caption{Decrease of the instantaneous Strehl ratio due to photon noise in dependence on the number of sensing elements $N$ and on the average number of photons, $Z$, received by the sensor during the acquisition of the two extra-focal images.
Other parameter values are: $l=0.05 \sqrt{N}$ [m], $D=8$\,m, $f=120$\,m and $\lambda=0.7\mu$m.}
\label{fig:strehl}
\end{center}
\end{figure}

As seen on Fig.\,\ref{fig:G}, the error propagation factors obtained with {\it Yao\/} are in good agreement with the values obtained in terms of the iterative method (sec.\,\ref{sec:nicolas} and \ref{sec:constant}). The fact that they do not coincide is probably due to the iterative method and {\it Yao\/} considering square and circular pupils respectively.
A general expression for the error propagation factor on a circular pupil is derived:
\begin{eqnarray}
G = \frac{\sigma^2_{\phi} }{\sigma_v^2} =  (\frac{l}{\sqrt{N}})^2\cdot(\frac{D}{f})^4 \cdot \frac{1}{\lambda^2} \cdot \left[ (0.64\pm 0.04) - \frac{(2.7\pm 0.3)}{N}\right]
\end{eqnarray}
The decrease in Strehl due to photon noise is then given by the following relation:
\begin{eqnarray}
\frac{S_1}{S_0} &=& \exp (- \sigma^2_{\phi}) =  \exp (-G\cdot \frac{N} {Z}) \\
\frac{S_1}{S_0} &=& \exp (- \frac{G_0}{Z} \cdot \left[ (0.64\pm 0.04)\,N - (2.7\pm 0.3) \right]  ) 
\end{eqnarray}
where $S_1$ and $S_0$ are the Strehl ratios with and in absence of photon noise. $G_0$ is defined by Eq.\,\ref{eq:G0},  $N$ is the number of sensing elements and $Z$ the total number of photons collected by the sensor during the acquisition of the two extra-focal images. 
Figure\,\ref{fig:strehl} traces the Strehl loss due to photon noise for an 8\,m telescope, in dependence on the number of sensing elements and on the photon number.

\section{Conclusion}

In this article, we have derived -- for any given curvature sensor -- a minimum required number of sub-apertures on the pupil-edge, in dependence on its total sub-aperture number.

We further examined the distribution of the sensor signal in terms of its mean, variance, skewness and kurtosis. The approximation in terms of a Gaussian distribution is found to be correct, even for fairly low photon numbers in the 50-100 range (mean photon number per sub-aperture and iteration). Yet, one requires much higher photon numbers (in the $10^3-10^4$ range) to obtain meaningful measurements at small signal values, i.e. at small wavefront phase distortions.

Finally, and in accordance with N.\,Roddier, the error propagation term has been shown to increase linearly with $N$ -- the number of sub-apertures -- if the sampling-surface stays constant. That is, if the pupil-surface likewise increases linearly with $N$. For a given telecope, however, and a fixed pupil-surface, the error propagation term is almost independent of $N$. This has been derived by an iterative method inspired by Hudgin's phase-reconstruction algorithm, and has been validated with an adaptive optics simulations software. A closed expression for the error propagation factor has been suggested.

\section*{Acknowledgments}
I thank Mark Chun and Christ Ftaclas for many helpful discussions, and I am grateful for the referees' constructive suggestions.  
Financial support from the United States Air Force Office of Scientific Research is likewise acknowledged.

\end{document}